\documentclass[conference,10pt]{IEEEtran}

\usepackage{amsmath, mathrsfs, mathtools}
\usepackage{amsfonts}
\usepackage{amssymb}
\usepackage{color}
\usepackage{multirow}
\usepackage{stmaryrd}
\usepackage{yfonts}
\usepackage{mathabx}
\allowdisplaybreaks
\usepackage{caption}
\usepackage{subcaption}
\usepackage{float}
\usepackage{stfloats}
\usepackage{amsfonts}
\usepackage{amssymb}
\usepackage{color}
\usepackage{multirow}
\usepackage{graphicx}
\usepackage{tcolorbox}
\usepackage{mathrsfs}
\usepackage[numbers,sort&compress]{natbib}
\usepackage{multicol}
\usepackage{algorithm}
\usepackage{algpseudocode}
\usepackage{pifont}
\usepackage{varwidth}
\usepackage{float}
\usepackage{afterpage}
\usepackage{balance}
\allowdisplaybreaks
\usepackage{lipsum}
\usepackage{mathbbol}

\def\mindex#1{\index{#1}}

\def\sq{\hbox{\rlap{$\sqcap$}$\sqcup$}}
\def\qed{\ifmmode\sq\else{\unskip\nobreak\hfil
\penalty50\hskip1em\null\nobreak\hfil\sq
\parfillskip=0pt\finalhyphendemerits=0\endgraf}\fi\medskip}

\long\def\defbox#1{\framebox[.9\hsize][c]{\parbox{.85\hsize}{%
\parindent=0pt
\baselineskip=12pt plus .1pt      
\parskip=6pt plus 1.5pt minus 1pt 
 #1}}}

\long\def\beginbox#1\endbox{\subsection*{}%
\hbox{\hspace{.05\hsize}\defbox{\medskip#1\bigskip}}%
\subsection*{}}

\def\endbox{}

\newsavebox{\junk}
\savebox{\junk}[1.6mm]{\hbox{$|\!|\!|$}}

\def\bC{{\mathbb C}}

\def\bE{{\mathbb E}}

\def\bR{{\mathbb R}}

\def\sfH{{\sf H}}

\def\bfmath#1{{\mathchoice{\mbox{\boldmath$#1$}}%
{\mbox{\boldmath$#1$}}%
{\mbox{\boldmath$\scriptstyle#1$}}%
{\mbox{\boldmath$\scriptscriptstyle#1$}}}}

\def\bfmY{\bfmath{Y}}

\def\bfmhhaY{\bfmath{\hhaY}} 
\def\bfmhhaY{\hbox to 0pt{$\widehat{\bfmY}$\hss}\widehat{\phantom{\raise 1.25pt\hbox{$\bfmY$}}}}

\def\til={{\widetilde =}}

\def\clC{{\cal C}}

\def\clN{{\cal N}}

 \def\FRAC#1#2#3{\genfrac{}{}{}{#1}{#2}{#3}}

\def\ddtp{{\mathchoice{\FRAC{1}{d^{\hbox to 2pt{\rm\tiny +\hss}}}{dt}}%
{\FRAC{1}{d^{\hbox to 2pt{\rm\tiny +\hss}}}{dt}}%
{\FRAC{3}{d^{\hbox to 2pt{\rm\tiny +\hss}}}{dt}}%
{\FRAC{3}{d^{\hbox to 2pt{\rm\tiny +\hss}}}{dt}}}}

\def\average#1,#2,{{1\over #2} \sum_{#1}^{#2}}

\def\eye(#1){{\bf(#1)}\quad}

\newtheorem{remark}{{\bf Remark}}

\def\eq#1/{(\ref{e:#1})}

\newcommand{\beqn}[1]{\notes{#1}%
\begin{eqnarray} \elabel{#1}}

\newcommand{\eeqn}{\end{eqnarray} }

\newcommand{\beq}[1]{\notes{#1}%
\begin{equation}\elabel{#1}}

\newcommand{\eeq}{\end{equation}}

\def\bdes{\begin{description}}
\def\edes{\end{description}}

\newcounter{rmnum}

\newcounter{anum}

{\end{list}}

\def\ass(#1:#2){(#1\ref{#1:#2})}

\def\ritem#1{
\item[{\sf \ass(\current_model:#1)}]
}

\newenvironment{recall-ass}[1]{%
\begin{description}
\def\current_model{#1}}{
\end{description}
}

\newcommand{\diag}{{\rm diag}}

\newcommand{\Sigmay}{{\Sigmay}_{\yv}}

\def\herm{{\sfH}}

\allowdisplaybreaks

\def\cg{{\clC\clN}} 
\newcommand{\normd}[1]{{\left\vert\kern-0.25ex\left\vert\kern-0.25ex\left\vert #1 
		\right\vert\kern-0.25ex\right\vert\kern-0.25ex\right\vert}}
\long\def\comment#1{}

\newcommand{\av}{{\bf a}}

\newcommand{\cv}{{\bf c}}

\newcommand{\gv}{{\bf g}}
\newcommand{\hv}{{\bf h}}

\newcommand{\sv}{{\bf s}}

\newcommand{\vv}{{\bf v}}
\newcommand{\xv}{{\bf x}}
\newcommand{\yv}{{\bf y}}
\newcommand{\zv}{{\bf z}}

\newcommand{\Bm}{{\bf B}}
\newcommand{\Cm}{{\bf C}}

\newcommand{\Fm}{{\bf F}}
\newcommand{\Gm}{{\bf G}}
\newcommand{\Hm}{{\bf H}}

\newcommand{\Wm}{{\bf W}}
\newcommand{\Vm}{{\bf V}}
\newcommand{\Xm}{{\bf X}}
\newcommand{\Ym}{{\bf Y}}

\newcommand{\Dc}{{\cal D}}

\newcommand{\Pc}{{\cal P}}

\newcommand{\gammav}{\hbox{\boldmath$\gamma$}}

\newcommand{\lambdav}{\hbox{\boldmath$\lambda$}}

\newcommand{\Sigmam}{\hbox{\boldmath$\Sigma$}}

\newcommand{\Thetam}{\hbox{\boldmath$\Theta$}}

\newcommand{\trace}{{\rm Tr}}

\makeatletter
\def\ps@IEEEtitlepagestyle{
  \def\@oddfoot{\mycopyrightnotice}
  \def\@evenfoot{}
}
\def\mycopyrightnotice{
  {\footnotesize This work has been submitted to the IEEE for possible publication. Copyright may be transferred without notice, after which this version may no longer be accessible.\hfill} 
  \gdef\mycopyrightnotice{}
}

\@ifundefined{showcaptionsetup}{}{
 \PassOptionsToPackage{caption=false}{subfig}}
\makeatother

\usepackage{eso-pic}
\newcommand\AtPageUpperMyright[1]{\AtPageUpperLeft{
 \put(\LenToUnit{0.5\paperwidth},\LenToUnit{-1cm}){
     \parbox{0.5\textwidth}{\raggedleft\fontsize{9}{11}\selectfont #1}}
 }}
\newcommand{\conf}[1]{
\AddToShipoutPictureBG*{
\AtPageUpperMyright{#1}
}
}

\title{Deep-Learning Aided Channel Training and Precoding in FDD Massive MIMO with \\Channel Statistics Knowledge}
\conf{Accepted by IEEE International Conference on Communications 2023}
\author{Yi Song, Tianyu Yang, Mahdi Barzegar Khalilsarai, and Giuseppe Caire\\
Technische Universit\"{a}t Berlin, Berlin, 10623, Germany.\\
Emails: $\{$yi.song, tianyu.yang, m.barzegarkhalilsarai, caire$\}$@tu-berlin.de

}

\begin{document}
	
\maketitle
	
\begin{abstract}
	We propose a method for channel training and precoding in FDD massive MIMO based on deep neural networks (DNNs), exploiting Downlink (DL) channel covariance knowledge. The DNN is optimized to maximize the DL multi-user sum-rate, by producing a pre-beamforming matrix based on user channel covariances that maps the original channel vectors to ``effective channels". Measurements of these effective channels are received at the users via common pilot transmission and sent back to the base station (BS) through analog feedback without further processing. The BS estimates the effective channels from received feedback and constructs a linear precoder by concatenating the optimized pre-beamforming matrix with a zero-forcing precoder over the effective channels. We show that the proposed method yields significantly higher sum-rates than the state-of-the-art DNN-based channel training and precoding scheme, especially in scenarios with small pilot and feedback size relative to the channel coherence block length. Unlike many works in the literature, our proposition does not involve deployment of a DNN at the user side, which typically comes at a high computational cost and parameter-transmission overhead on the system, and is therefore considerably more practical. 
\end{abstract}
	
\section{Introduction}
Deep Neural Networks (DNNs) have been recently successfully applied in various areas of wireless communications such as resource allocation and scheduling \cite{eisen2020optimal, cui2019spatial}, channel estimation \cite{balevi2020massive,mashhadi2021pruning}, beamforming \cite{hojatian2021unsupervised}, transceiver design \cite{honkala2021deeprx}, etc. Given sufficient data, a DNN is trained in a (semi-)supervised or unsupervised fashion to learn mappings from an input space to some desired output that optimizes a suitable utility metric that is otherwise very hard to optimize with conventional tools. 
	
In this paper, we propose a DNN-based solution to the problem of channel training and multi-user precoding in a frequency division duplex (FDD) massive MIMO system with channel statistics knowledge at the Base Station (BS). It is well-known that, to achieve the benefits of massive MIMO, the transmitter needs to obtain fresh Downlink (DL) channel state information (CSI). Unlike time division duplexing (TDD) systems, where relying on channel reciprocity, DL channels are directly estimated from Uplink (UL) pilots, in FDD the BS must train the CSI by broadcasting pilot sequences in DL and receiving user feedback. This process requires careful design of DL pilots, user feedback messages, and the precoder based on feedback. In particular, a small pilot length (in DL) and feedback size (in UL) relative to the channel dimension, results in poor DL spectral efficiency. This is caused by the large channel estimation error and the resulting interference due to precoding over erroneous channels. Incorporating knowledge of channel statistics at the transmitter in designing the pilots and the precoder can significantly mitigate this effect. We propose a scheme in which a DNN is trained with the given constraints on pilot and feedback size (fixed by the standard) to produce a pre-beamforming matrix as a function of user channel statistics. Both the pilot vectors (a set of $T_{\rm dl}$ row-vectors in $\bC^M$) and precoding vectors (a set of $K$ row-vectors in $\bC^M$) are chosen from the row-space of this pre-beamforming matrix. As will be apparent by the signal model in the next sections, the pre-beamforming matrix maps original channels to ``effective channels", which will be estimated (through DL training and UL feedback) and over which zero-forcing (ZF) will be performed. Intuitively, since an ``accurate" estimation of the original channels with limited pilot and feedback resources is infeasible, the DNN-based transform is employed to manage interference by precoding over certain effective channels with possibly smaller number of known coefficients and therefore can be trained with the given pilot/feedback budget. Other elements of our proposed network include the following. Upon receiving pilots, the users send them back to the BS after a power normalization via analog feedback \cite{marzetta2006fast}, i.e., feeding back complex-valued measurements by modulating them as quadrature and in-phase components of the baseband signal. The BS then computes a minimum mean-squared error (MMSE) estimate of the effective channels. The precoder is then generated as a product of a ZF precoder on the effective channels and the pre-beamforming matrix and is used to send data to users in DL. The DNN is optimized end-to-end to yield a pre-beamforming matrix based on input channel statistics, that maximizes the multi-user sum-rate.

Recently many works have utilized DNNs for channel training and precoding in massive MIMO. Some have proposed extrapolation of DL channels from UL channels using DNNs \cite{alrabeiah2019deep,arnold2019towards,yang2019deep}. Albeit highly successful under certain scenarios, these methods would fail when the channel coherence bandwidth is small relative to the separation between UL and DL carrier frequencies and explicit DL training and feedback is necessary. In other works, pilot design and channel estimation with DNNs is considered \cite{mashhadi2021pruning, ma2020data}. The objective in these works is to minimize the channel estimation MSE, which is different from maximizing the multi-user sum-rate as considered in our work. Another category of works focuses on compression of feedback, where perfect channel state knowledge at the users is assumed \cite{mashhadi2020distributed,wen2018deep,guo2020deep}. This assumption is hard to achieve in massive MIMO, since the channel dimension is typically larger than the pilot length and channel estimation is carried out via a compressed sensing scheme which not only may fail depending on the channel sparsity order, but is also computationally costly and difficult to implement in real time in the user devices. Finally, \cite{sohrabi2021deep} proposed a highly successful DNN-based scheme for pilot sequence design, feedback quantization and DL precoding. This scheme however involves deployment of the feedback computation layers at the user side, which requires transfer of a large (in the order of a million) number of parameters to the users, incurring a huge overhead in DL. 

Our proposed method offers the following advantages with respect to the existing works in the literature.
\begin{enumerate}
	\item \textbf{Exploiting Channel Statistics:} Our proposed DNN utilizes channel statistics to design DL pilots and the precoder. This results in significantly higher DL sum-rate, particularly in scenarios where channel training is difficult due to small pilot and feedback dimensions. Note that, although availability of channel statistics knowledge at the BS is not always granted, it is justified by the fact that in FDD systems, DL channel covariances can be estimated from UL pilots based on what is known as ``angle reciprocity" \cite{xie2016unified,miretti2018fdd, haghighatshoar2018multi}. Therefore, it is reasonable to devise DNN-based solutions that exploit channel covariance knowledge.
		
	\item \textbf{No DNN at the User Side:} Unlike many works in the literature \cite{mashhadi2020distributed,wen2018deep,guo2020deep,sohrabi2021deep}, our proposed method does not involve training a DNN at the user side or transmission of optimized DNN parameters to it. Users simply send back pilot measurements to the BS with a power normalization, from which the BS estimates effective channels.
	
	\item \textbf{Direct Sum-Rate Maximization:} 
	The idea of training and precoding in FDD massive MIMO by pre-conditioning channels with a transform based on statistics was proposed by some of the authors of the present work in \cite{khalilsarai2018fdd}. There, the transform was optimized to maximize the spatial multiplexing gain, which is equivalent to the rate pre-log factor in high SNR. In contrast, in the present work we optimize the transform to directly maximize ergodic sum-rate in a data-driven fashion and through the DNN. Our simulation results will show that this approach significantly improves upon \cite{khalilsarai2018fdd}.
\end{enumerate}

We will show via simulations that our method achieves better performance in terms of DL sum-rate compared to the state-of-art result in \cite{sohrabi2021deep} as well as \cite{khalilsarai2018fdd}, especially when the pilot and feedback dimensions are small compared to the channel dimension, proving the applicability of this scheme in FDD massive MIMO systems.  

\section{System model}
\subsection{Common Training}
We consider a massive MIMO system in FDD mode, where a BS equipped with a uniform linear array (ULA) of $M$ antennas servers $K$ users with a single antenna in a cell. Because channel reciprocity does not hold in FDD, the BS has to train DL channels by broadcasting pilot sequences of length $\beta$ from each of its $M$ antenna ports. We denote these pilot sequences as rows of a pilot matrix $\Xm^{\rm p}\in \bC^{\beta \times M} $ (the superscript ``p" stands for ``pilot"). The pilot signal received at user $k$ is expressed as
\begin{equation}\label{eq:dl_pilots}
	\yv_k^{\rm p} =  \Xm^{\rm p}\hv_k + \zv_k^{\rm p}, \; k\in[K],
\end{equation}
where $\hv_k \sim\mathcal{CN}(\mathbf{0},\Cm_k)$ is the Rayleigh fading channel vector of user $k$ with covariance $\Cm_k=\bE[\hv_k \hv_k^\herm]$, $\zv_k^{\rm p}\sim \cg (\mathbf{0},\mathbf{I}_M)$ is additive white Gaussian noise (AWGN) with unit variance per element, and for an integer $a$ we define $[a]\triangleq \{1, 2, \dots, a\}$. Assuming the BS has a total transmission power of $P_{\rm dl}$, the pilot matrix should satisfy the power constraint
\begin{equation}\label{eq:power_tr}
    \|[\Xm^{\rm p}]_{i,\cdot}\|^2 \leq P_{\rm dl}, \; \forall i \in [\beta].
\end{equation}
where $[\Xm^{\rm p}]_{i,\cdot}$ denotes the $i$-th row of $\Xm^{\rm p}$. Also, since the noise variance is normalized to one, we define the SNR in DL as $\text{SNR}_{\rm dl}= P_{\rm dl}$. For future reference, we define the effective channel of user $k$ by
\begin{equation}\label{eq:eff_channels}
    \gv_k \triangleq \Bm \hv_k, \;\forall k \in [K],
\end{equation}
where $\Bm\in \bC^{M\times M}$ is the pre-beamforming matrix, mapping the original channel to the effective channel.

We propose to design $\Xm^{\rm p}$ as the product 
\begin{equation}\label{eq:X_decomp}
    \Xm^{\rm p} = \Wm \Bm, 
\end{equation}
where $\Wm \in\bC^{\beta \times M}$ is a an arbitrary full-rank matrix. While we do not impose any constraints on $\Wm$ other than being full-rank, $\Bm$ will be produced by a trained DNN with user channel covariances as input. This will be explained in Section~\ref{sec:DNN_Based_Design}. With this construction, the DL pilot signal in \eqref{eq:dl_pilots} can be equivalently written as $\yv_k^{\rm p} =  \Wm \gv_k + \zv_k^{\rm p}$, so that received pilot symbols can be equivalently seen as noisy linear measurements of the effective channel. 

\subsection{Analog Feedback}
After receiving pilot signals, each user sends a feedback ``message" to the BS using the UL channel. A common approach known as digital feedback consists of estimating the channel at the receiver from the pilots, quantizing it and sending the quantization index to the BS \cite{caire2010multiuser}. Alternatively, users can encode the pilot signal into quantization codewords without explicit channel estimation. A different approach, known as analog feedback consists of sending complex-valued feedback symbols to the BS by modulating the quadrature and in-phase components of the carrier by real and imaginary parts of the feedback symbol \cite{marzetta2006fast}. Analog feedback is simpler and imposes less feedback delay than digital feedback which requires quantization and channel coding. The feedback symbols can be estimates of the channel or the received pilot signal itself. In our proposition, the user sends the power-normalized pilot symbols directly and without channel estimation to the BS via analog feedback. The feedback message of user $k$ in this case is given by
\begin{equation}
    \xv^{\rm fb}_k = \sqrt{\rho_k} \yv^{\rm p}_k,~~ \text{with}~ \rho_k = \beta P_{\rm ul}/\|\yv^{\rm p}_k\|^2.  
\end{equation}
which satisfies the average power constraint
\begin{equation}\label{eq:power_fb}
    \|\xv^{\rm fb}_k\|^2 \leq \beta P_{\rm ul}, \;\forall k \in [K],
\end{equation}
where $P_{\rm ul}$ is the user average transmit power per symbol in UL, assumed equal among all users. 
The elements of $\xv^{\rm fb}_k$ are sent back to the BS via analog feedback. This means that, just like the analog QAM modulation, real and imaginary parts of each complex-valued symbol in the feedback message modulate carriers that have a 90 degrees phase difference. These carriers are then combined and the resulting signal is sent to the BS. To avoid any confusion, we emphasize that the feedback symbols are not quantized and the user does \textit{not} use a digital QAM modulation here. Also, note that there is no factual problem with transmitting unquantized feedback symbols: even in the prevalent OFDM signaling with digital QAM, the continuous time-domain I and Q signal after the IFFT is transmitted effectively unquantized (quantized with 10-12 bits per sample). Besides, we model the UL channel as an AWGN channel which is orthogonally accessed by the users. Then, the BS receives the noisy feedback signal as 
\begin{align}
    \yv^{\rm fb}_k = \xv^{\rm fb}_k  + \zv^{\rm fb}_k,
\end{align}
where $\zv^{\rm fb}_k \sim \mathcal{CN}(\mathbf{0}, \mathbf{I}_{\beta})$ is the noise vector.

\begin{remark}
    The UL channel can be generally modeled as a multiple-access channel (MAC), but we consider the special case of the AWGN channel with orthogonal access for simplicity and defer the treatment of more general models to a future work. Note that most previous works do not discuss the feedback channel model at all and assume availability of perfect, error-free feedback at the BS \cite{wen2018deep,lu2020multi,sohrabi2021deep}. \hfill $\lozenge$
\end{remark}
\begin{remark}
    By adopting the proposed analog feedback strategy, there is no need for complex processing at the user side. This is in contrast to schemes that involve deploying a DNN at the user side (see e.g., \cite{sohrabi2021deep,mashhadi2021pruning,guo2020deep}) that have two disadvantages: First, the forward pass of a DNN involves consecutive matrix multiplications and applying activation functions which consume time. Second, and more importantly, if the DNN is trained at the BS side, its optimized parameters should be transferred to the user as soon as it enters the cell. Given that the number of parameters in a DNN can be in the order of millions, this imposes a large overhead on DL resources. Our scheme avoids both of these disadvantages by using analog feedback. \hfill $\lozenge$
\end{remark}

\subsection{Effective Channel Estimation and Precoding}
Given the feedback signal, the BS computes an MMSE estimate of effective channels as
\begin{equation}\label{eq:est_g}
	\widehat{\gv}_k = \bE [\gv_k | \yv_k^{\rm fb}] = \Cm_{gy, k} \Cm_{yy, k}^{-1} \yv^{\rm fb}_k,~k\in [K],
\end{equation}
where 
\begin{align}
	\Cm_{gy, k} &= \mathbb{E} [\gv_k (\yv^{\rm fb}_k)^\herm]= \sqrt{\rho_k} \Bm \Cm_k(\Xm^{\rm p})^\herm,\\
	\Cm_{yy,k} &= \mathbb{E} [\yv^{\rm fb}_k(\yv^{\rm fb}_k)^\herm]  = \rho_k \Xm^{\rm p} \Bm \Cm_k\Bm^\herm (\Xm^{\rm p})^\herm + (1 + \rho_k) \mathbf{I}_{\beta}.
\end{align}
Next, the BS transmits data in DL using a linear precoder as follows. Let $\sv = [s_1,\ldots, s_K]$ denote a row vector consisting of the user data symbols, each satisfying $\bE [|s_k|^2] = 1$. The precoded data vector is then given by	$\xv^{\rm d} = \sv \Vm\in\,  \bC^{1\times M}$,
where $\Vm$ is a linear precoding matrix and the superscript ``d" stands for ``data". Similar to the design of the pilot matrix in \eqref{eq:X_decomp}, we propose a construction of the precoder as the product 
\begin{equation}\label{eq:precoder}
    \Vm = \widetilde{\Vm} \Bm, 
\end{equation}
where $\Bm$ is the pre-beamforming matrix to be designed and $\widetilde{\Vm}$ is a zero-forcing precoder on the estimated effective channels. Denoting the estimated effective channels by $\widehat{\Gm} = \left[ \widehat{\gv}_1,\dots,\widehat{\gv}_K \right]$, this precoder is given by
\begin{equation}\label{eq:precoder_tilde}
	\widetilde{\Vm} = \sqrt{\alpha} \left( \widehat{\Gm}^\herm \widehat{\Gm} \right)^{-1} \widehat{\Gm}^\herm \, \in\,  \bC^{K\times M},
\end{equation}
where each row represents the precoding vector of a user and $\alpha>0$ is a scalar that forces the precoder to satisfy
\begin{equation}\label{eq:power_V}
	\trace(\Vm\Vm^\herm) \leq P_{\rm dl}.
\end{equation}
The point of decomposing the precoder in \eqref{eq:precoder} is for it to first map the original channel to the effective channel through $\Bm$ and then apply zero-forcing on the effective channels. The received data symbol at user $k$ is given as 
\begin{align}
    y^{\rm d}_k &= \xv^{\rm d} \hv_k + z^{\rm d}_k \\
	            & = \vv_k \hv_k  s_k + \sum^K_{k'\neq k} \vv_{k'} \hv_k s_{k'} + z_k^{\rm d},
\end{align}
where $\vv_k$ is the $k$-th row of $\Vm$ and $z_k^{\rm d}\sim \cg (0,1)$ is the AWGN. Treating interference as noise and assuming signal and interference coefficients knowledge at the receiver, the achievable ergodic sum-rate in DL is given by \cite{caire2018ergodic}
\begin{align}\label{eq:ergodic_sum_rate}
    R_{\rm sum} &= \sum_{k=1}^K\bE\left[ \log_2\left(1+\frac{|\vv_k\hv_k|^2}{1 + \sum_{k'\neq k} |\vv_{k'}\hv_k|^2}\right)\right],
\end{align}
where the expectation is taken over channel and noise distributions. Note that the terms $\{|\vv_{k'}\hv_k|^2~:~k,k'\in [K],\, k\neq k' \}$ are interference coefficients between channels of users $k$ and $k'$. The precoder $\Vm$ is a function of the pre-beamforming matrix $\Bm$ through the pilot matrix $\Xm^{\rm p}$ in \eqref{eq:X_decomp}, the channel estimates in \eqref{eq:est_g} and the resulting precoder in \eqref{eq:precoder}. Our goal is to design $\Bm$ based on available DL channel covariances at the BS, such that the ergodic sum-rate is maximized, i.e., we want to find a mapping $\Pc_{\Bm}(\cdot)$ from the set of $K$ user channel covariances to the pre-beamforming matrix $\Bm$ that maximizes ergodic sum-rate. We further denote the mapping described by Eqs. \eqref{eq:est_g}-\eqref{eq:power_V} from user feedback signals denoted by $\Ym^{\rm fb} = \left[ \yv_1^{\rm fb},\ldots, \yv_K^{\rm fb}  \right]$ to the precoder by $f_{\rm pc}\left(\,\cdot\,;\Bm\right)$, so that $\Vm = f_{\rm pc}\left(\Ym^{\rm fb};\Bm\right)$. Now, the sum-rate maximization problem can be posed as:
\begin{subequations}\label{eq:sum_rate_max}
\begin{align}
    &\underset{\Pc_{\Bm}(\cdot)}{\text{maximize}} \quad && R_{\rm sum} \\
    & \text{subject to}  && \Bm = \Pc_{\Bm} \left(\left\{\Cm_k \right\}_{k=1}^K \right),\\
	&~&&\Vm = f_{\rm pc}(\Ym^{\rm fb}; \Bm), \label{eq:f_pc} \\
	& ~ &&\|[\Wm \Bm]_{i,\cdot}\|^2 \leq P_{\rm dl}, \; \forall i \in [\beta], \label{eq:power_WB}\\
	 &~&&\eqref{eq:power_fb}, \eqref{eq:power_V}.
\end{align}
\end{subequations}

\section{Pre-Beamforming Based on Channel Statistics}\label{sec:DNN_Based_Design}
In massive MIMO, it is typical to have a pilot length that is small relative to the channel dimension ($\beta<M$). This results, from \eqref{eq:dl_pilots} in an underdetermined system of noisy linear equations
from which the effective channel $\gv_k = \Bm \hv_k$ must be estimated. Because the system is underdetermined, the channel estimation error can be high even with the MMSE estimator. Given the effective channel covariance $\bE [\gv_k \gv_k^\herm] = \Bm \Cm_k \Bm^\herm$, it is shown in  \cite{khalilsarai2018fdd} that if $\beta <\text{rank}\left( \Bm \Cm_k \Bm^\herm \right)$, then the effective channel estimation MSE scales as $O(1)$ when $\text{SNR}_{\rm dl}\to \infty$. This means that a small pilot dimension leads to a constant channel estimation error which is independent of SNR. In addition, when the channel estimation error is large, naive zero-forcing results in large interference coefficients between the users in the denominator of \eqref{eq:ergodic_sum_rate} and reduces the ergodic sum-rate. Thus, the pre-beamforming matrix should be designed such that the rank of the effective channel covariance $\Bm \Cm_k \Bm^\herm$ becomes smaller to reduce the estimation error with a given pilot dimension. On the other hand, the effective rank should not reduce too much because then the signal coefficient $|\vv_k \hv_k|^2$ in the numerator of \eqref{eq:ergodic_sum_rate} reduce, resulting in smaller ergodic sum-rate. In the extreme case, if $\Bm=\mathbf{0}$ then the sum-rate will be zero. These two effects imply that the pre-beamforming matrix $\Bm$ should be a transformation that reduces the inherent dimension (i.e., channel covariance rank) of effective channels down to a certain value to achieve a favourable trade-off in minimizing interference and maximizing signal coefficients. 
	
We simplify the design problem by exploiting properties of the channel covariance. It is known that the covariance of a ULA channel with large $M$ is (approximately) diagonalized by the DFT matrix, thanks to the similarity of large Toeplitz matrices to their Circulant equivalents and the famous Szegö's theorem \cite{adhikary2013joint, zhu2017asymptotic}. In other words, the channel covariance can be approximately decomposed as
\begin{equation}
	\Cm_k\approx \Fm \text{diag}(\gammav_k) \Fm^\herm,
\end{equation}
where $\gammav_k \in \bR_+^M$ is the vector of channel covariance eigenvalues of user $k$ and $\Fm \in \mathbb{C}^{M\times M}$ is the DFT matrix whose $(m, n)$-th entry is given by $[\Fm]_{m, n} = \frac{1}{\sqrt{M}} e^{-j 2 \pi \frac{m n}{M}}, m, n \in [M]$. 
We simplify design of $\Bm$ by restricting it to belong to the set 
\begin{equation}
    \mathcal{B} \triangleq\{\diag(\lambdav)\Fm^\herm\; : \;\lambdav\in[0,1]^M\}.
\end{equation}
Then the covariance of the effective channel is given by
\begin{equation}\label{eq:cov_eff}
	\Bm \Cm_k \Bm^\herm \approx \text{diag}\left( \lambdav^2 \odot \gammav_k \right)
\end{equation}
where $\lambdav^2$ denotes element-wise square of $\lambdav$ and 
$\odot$ denotes element-wise product. Essentially with this design choice, the ``effective rank" of the covariance is equivalent to the number of large coefficients in $\lambdav$ and therefore this vector controls the inherent dimension of effective channels. From a different perspective, $\lambdav$ can be seen as ``beam-selection" vector, since the DFT columns are equivalent to the array steering vectors of a ULA evaluated on a grid of angle-of-departures (AoDs). If the $m$-th coordinate of $\lambdav$ ($\lambda_m\in [0,1]$) is small ($\lambda_m\to 0$), then the contribution of the $m$-th beam in the effective channel of all users will be eliminated. In this sense, the present work is similar to the \textit{active channel sparsification} (ACS) method in \cite{khalilsarai2018fdd} which proposed beam-selection with the objective of maximizing  the multiplexing gain. However, the DNN-based method proposed here aims to directly maximize the sum-rate and in this sense extends the idea presented in \cite{khalilsarai2018fdd}.

\subsection{DNN-Based Optimization}
\begin{figure*}[ht] 
    \centering
    \includegraphics[width=0.9\textwidth]{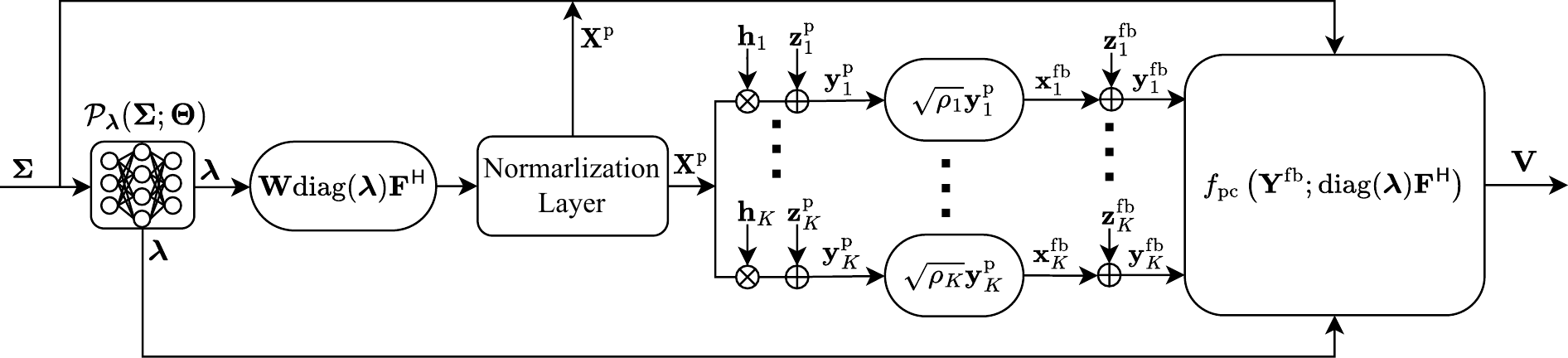}
    \caption{System schematic for DNN aided FDD multi-user training and precoding with channel statistics knowledge. The proposed system takes DL covariance matrices $\Sigmam$ as input and train with DL channels $\{\hv_k\}$ to output the precoder $\Vm$.}
    \label{fig:DNN}
    \vspace{-3mm}
\end{figure*}

We employ a DNN to produce $\lambdav$ vector based on input channel covariances $\{ \Cm_k \}_{k=1}^K$. Based on the $K$ covariances, the given pilot dimension and SNR, the network is trained to output a $\lambdav$ that maximizes the sum-rate. Note that the channel covariance of a ULA is a Toeplitz Hermitian matrix that is fully determined by its first column. Denoting the first columns of the $K$ covariances by $\cv_1,\ldots,\cv_K$, we define the matrix $\Sigmam = \left[\cv_1,\ldots,\cv_K \right]\in \bC^{M\times K}$. Then the optimization problem \eqref{eq:sum_rate_max} can be reformulated as
\begin{subequations}\label{eq:sum_rate_max_DNN}
\begin{align}
    &\underset{{\Thetam}}{\text{maximize}} \; && R_{\rm sum} \label{eq:obj_DNN}\\
 	& \text{subject to}  && \lambdav= \Pc_{\lambdav}\left(\Sigmam;\Thetam\right),\\
 	&~&& \Bm = \diag(\lambdav)\Fm^\herm, \\
 	&~&&\eqref{eq:f_pc}, \eqref{eq:power_WB}, \eqref{eq:power_fb}, \eqref{eq:power_V}, \label{subeq:constraint_power}
     \end{align}
\end{subequations}
where $\Pc_{\lambdav}(\cdot;\Thetam): \bC^{M\times K}\to \bC^M$ is the mapping from the covariance first columns to the beam-selection vector associated with the DNN with parameters $\Thetam$. The proposed architecture is illustrated in Fig.~\ref{fig:DNN}.

We solve \eqref{eq:sum_rate_max_DNN} in a data-driven fashion to optimize network parameters by generating random realizations of $\Sigmam$ according to a distribution $\Dc$. This distribution is typically based on geometric properties of the scattering environment, such as the number of paths, the distribution of AoDs and their associated powers. In practice, random realizations of this distribution can be collected at the BS at different times for $K$ randomly located users in the cell. In our simulation results, we consider random samples of $\Dc$ to be generated from a multipath scattering model that is independent across users and is parametrized by the number of paths, uniformly distributed AoDs and powers. Then, each random sample of $\Dc$ is given as input to the DNN. The expected value in the objective function \eqref{eq:obj_DNN} is replaced by an empirical mean obtained by generating many independent samples of the DL channel for each user.
\begin{remark}
The main difference between our design and the DNN-based scheme in \cite{sohrabi2021deep} is that we learn a mapping between $\Sigmam$ and $\Bm$, exploiting the channel second order statistics for different users. In contrast, \cite{sohrabi2021deep} proposes to learn a pilot matrix that should  ``fit'' all the user channel statistics from a large ensemble, and not the specific statistics of the $K$ users that are scheduled to be served in a single frame. \hfill $\lozenge$
\end{remark} 

\subsection{DNN Implementation Details}
Our DNN consists of three fully-connected layers, where the number of hidden neurons per layer are $[\ell_1, \ell_2, \ell_3] = [1024, 512, M]$. We use ReLU activation functions in all hidden layers. In order to produce $\lambdav$ in $[0,1]^M$, we use $tanh$ activation in the output layer and scale its output to $[0,1]$ as $0.5 (\tanh(\cdot) + 1)$. We implement the network in PyTorch \cite{paszke2019pytorch} with the Adam optimizer \cite{kingma2015adam} with a batch size of $1024$ and initial learning rate of $10^{-4}$. For fast convergence, a batch normalization layer is added before each linear layer \cite{ioffe2015BN}.

\section{Numerical Results}
For the simulations, we consider $M = 64$ antennas, $K = 6$ users, $\beta =8$ pilots. We stress the point that in general, estimating a set of $6$, $64$-dimensional channels from a DL pilot of length $8$ is extremely difficult and a successful performance in such a setup should be noticed. We set the DL SNR to $P_{\rm dl} = 20$ and consider a scattering channel model with $L$ paths. The DL channel covariance of user $k$ is given by
\begin{equation}
    \Cm_k = \sum_{\ell= 1}^{L}\eta_{k,\ell}\av(\theta_{k,\ell})\av^\herm(\theta_{k,\ell}),   \;\forall k \in [K],
\end{equation}
where $\eta_{k,\ell}$ and $\theta_{k,\ell}$ are the power and the AoD of the $\ell$-th channel path of user $k$, and where $\av(\theta)\in \bC^{M}$ is the steering vector of a ULA, whose $m$-th entry is given by $[\av(\theta)]_m = e^{j\frac{2\pi d}{\lambda'}(m-1)\sin(\theta)}$,  $m\in [M]$ where $d$ is the antenna spacing and $\lambda'$ is the carrier wavelength. We assume the maximum array angular aperture to be given by $\theta_{\rm max} = 60^\circ$ and assume that the antenna spacing is set to $d=\frac{\lambda'}{2\sin \theta_{\rm max}}$. The user AoDs are generated independently from a uniform distribution, i.e., $\theta_{k,\ell}\sim\mathcal{U}(-\theta_{\rm max},\theta_{\rm max})$. The path powers are randomly and uniformly generated in the real interval $[0.4, 0.8]$ and then scaled to sum to one, i.e., $\sum^{L}_{\ell=1} \eta_{k,\ell} = 1,\forall k\in[K]$. Choosing powers as such is not necessary, and is simply to avoid path powers close to zero. We recall that the input of the DNN is the matrix $\Sigmam$ containing the first covariance columns of all users as its columns. When generated according to the distribution of AoDs and powers as above, $\Sigmam$ follows a distribution $\Dc(L)$, parameterized by the number of paths $L$. This specific characterization is just used here to perform the simulations. In general, one can choose any family of distributions to generate $\Sigmam$ and train the DNN accordingly. In the upcoming simulations, we provide results for two important scenarios: (a) sparse scattering with $L=2$ paths per user, and (b) rich scattering with $L=20$ paths. Note that the number of paths is equivalent to the channel covariance rank. Given that the pilot length is $\beta=8$, training channels with a large $L$ is more difficult than those with a small $L$. For each case, we generate the training and testing data with randomly generated samples of $\Sigmam$ according to $\Dc (L)$. The training data is  per epoch randomly generated with a fixed series of random seeds.  The testing data contains $1000$ randomly samples of $\Sigmam$, and for each random sample of covariance, we generate $10$ random instantaneous channel samples as well as DL and UL additive noise vectors. The same testing data is used to produce results for all the baseline methods.

\subsection{Comparison Baselines}
We compare our proposed scheme with the state-of-the-art DNN-based design in \cite{sohrabi2021deep} that is under digital feedback and without channel statistic knowledge.\footnote{We have trained the DNN proposed in \cite{sohrabi2021deep} according to their public code. Our code can be found in https://github.com/YiSongTUBerlin/DL-Aided-Channel-Training-and-Precoding-in-FDD-Massive-MIMO-with-Channel-Statistics-Knowledge.git} The number of feedback symbols in analog feedback is $\beta$. Considering a UL channel capacity of $C_{\rm ul}=\log_2(1+P_{\rm ul})$ bits per channel use, this translates to $B=\beta C_{\rm ul}$ feedback bits. In order to make a fair comparison  between analog and digital feedback, we set the UL transmit power to $P_{\rm ul} = 2^{B/\beta}-1$ so that both strategies feed back the same amount of data. Additionally, we provide results of maximum ratio transmission (MRT) precoding and ZF precoding under perfect DL CSI. The precoder of MRT and ZF are respectively obtained by $\Vm_{\rm MRT} = J_{\rm MRT} \Hm^\herm$ and $\Vm_{\rm ZF} =J_{\rm ZF}  (\Hm^{\herm} \Hm)^{-1}\Hm^\herm$, where $\Hm = [\hv_1,\dots,\hv_K]$ and $J_{\rm MRT}$ and $J_{\rm ZF}$ are power normalization scalars to satisfy the power constraint \eqref{eq:power_V}. Furthermore, we also provide results for the case of training and precoding without the pre-beamforming matrix. This is equivalent to setting $\Bm = \Fm^\herm$ which performs only a rotation on the channel and is the same as setting $\lambdav = \mathbf{1}$ (a vector of all ones). Comparing to this case, the performance improvement by optimizing $\lambdav$ will become clear. Finally, both with and without pre-beamforming, the matrix constituent $\Wm$ of the pilot matrix in \eqref{eq:X_decomp} is generated randomly with standard normal elements and will be fixed in training. We noted earlier that the choice of this matrix is arbitrary as long as it is full-rank (which is the case, with probability 1, when each element generated as a standard normal random variable). We have tried optimizing this matrix, jointly with the rest of the network, but this did not result in noticeable gain in performance and therefore was ignored.
\subsection{Performance Comparison}
\begin{figure}
	\begin{center}
		\includegraphics[trim={0 5 0 10},width=0.9\columnwidth]{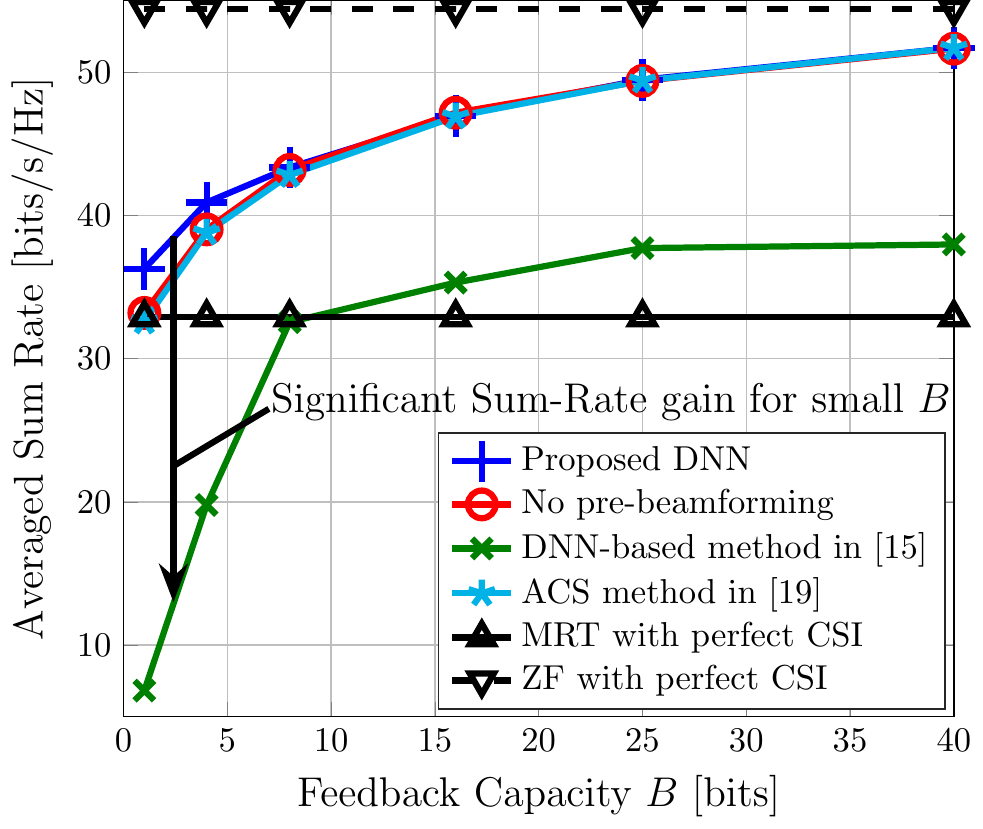}
		\caption{Sum-rate v.s. feedback capacity $B$ with $L=2$}
		\label{fig:results_L2}
	\end{center}
	\vspace{-5mm}
\end{figure}

 The sum-rate performance vs feedback capacity (in bits) for sparse scattering with $L=2$ is illustrated in Fig. \ref{fig:results_L2}. We observe that our proposed DNN-based technique outperforms all rival methods (except ZF with perfect CSI). In particular, we see a significant performance advantage in comparison to the DNN-based method in \cite{sohrabi2021deep}, especially for small feedback sizes. This should be mainly attributed to the fact that our proposed scheme exploits channel statistics knowledge at the BS. Even when there is practically no feedback ($B\to 0$), our scheme is able to achieve a relatively large sum-rate because one component of the designed precoder in \eqref{eq:precoder}, namely the pre-beamforming matrix $\Bm$ depends only on channel statistics and not the feedback. In this case, our DNN is essentially performing a kind of statistical beamforming with (almost) no CSI. Interestingly, statistical beamforming is shown to be very effective in the case of sparse channels \cite{liu2020statistical}, which supports the observed behavior here. The proposed method also outperforms MRT, which is due to the use of ZF precoding in our architecture. The advantage with respect to the case with no pre-beamforming matrix (the red curve) is rather small. This is due to the fact that the channels are sparse and $L<\beta$, and therefore with sufficient feedback no beam-selection is necessary. Here, the optimized beam-selection vector is $\lambdav\approx \mathbf{1}$, which is very close to no pre-beamforming with $\Bm = \Fm^\herm$. Finally, we see a similar advantage in comparison to the ACS method in \cite{khalilsarai2018fdd}, resulting from the direct maximization of sum-rate rather than multiplexing gain.
 
\begin{figure}
	\begin{center}
T		\includegraphics[trim={0 5 0 0},width=0.9 \columnwidth]{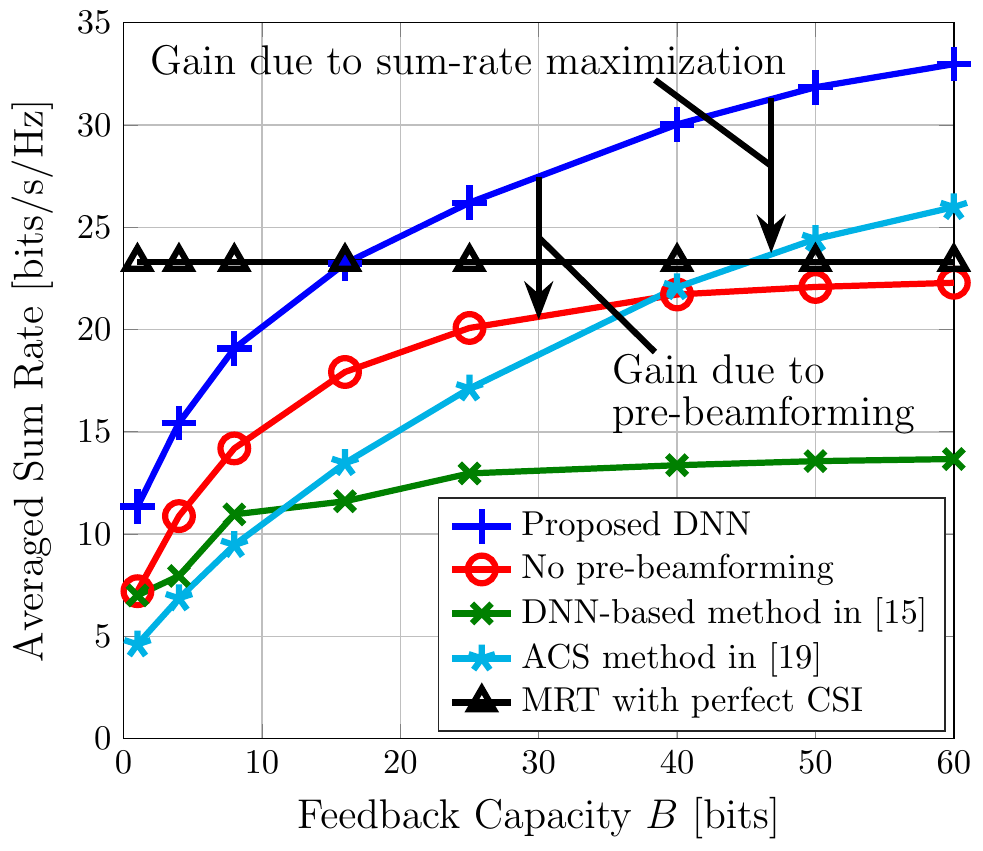}
		\caption{Sum-rate v.s. feedback capacity $B$ with $L=20$}
		\label{fig:results_L20}
	\end{center}
	\vspace{-7mm}
\end{figure} 

 The methods are compared in the rich scattering scenario with $L=20$ in Fig. \ref{fig:results_L20}. Here since $\beta<L$, we expect that using an optimized pre-beamforming matrix yields a large performance gain. This is confirmed by the results of Fig. \ref{fig:results_L20}, where we see that the proposed DNN-based method clearly outperforms the competitor methods. The performance advantage with respect to the case with no pre-beamforming is clearly observed and shows that even with channel statistics knowledge, the system under-performs because of the interference cause by MMSE channel estimation with $\beta<L$. The propsed method also performs better than the ACS method in \cite{khalilsarai2018fdd}, since it directly maximizes sum-rate instead of multiplexing gain. Finally, both channel statistics knowledge and optimized pre-beamforming  results in the much higher sum-rate values achieved by our method and the DNN-based method proposed in \cite{sohrabi2021deep} for all feedback sizes. We point out that ZF precoding with perfect CSI yields a sum-rate of $\approx60$ bits/s/Hz, which is much larger than the rest and is omitted from Fig. \ref{fig:results_L20} for a better representation of the results.

In Fig.~\ref{fig:lambda}, we present a heat map of the optimized $\lambdav$ of the proposed scheme for $50$ random realizations of $\Dc (L)$ (stacked as rows), for three different combinations of parameters, namely $(L=2, B=1)$ (sparse scattering, small feedback) in Fig.~\ref{fig:lambda_B1_L2}, $(L=20, B=40)$ (rich scattering, large feedback) in Fig.~\ref{fig:lambda_B40_L20}, and $(L=20, B=1)$ (rich scattering, small feedback) in Fig.~\ref{fig:lambda_B1_L20}. First, we observe in Fig.~\ref{fig:lambda_B1_L2} that under $L=2$ the learned $\lambdav$ are almost all ones because the channels are sparse enough that no pre-beamforming is needed and agrees with the sum-rate performance presented in  Fig.~\ref{fig:results_L2}. In Fig.~\ref{fig:lambda_B40_L20}, since $\beta<L$, the DNN produces beam-selection vectors that contain many zeros, meaning that many beams are not selected in the pre-beamformer. If we decrease the feedback size from $B=40$ to $B=1$ bits, we have the results of Fig. \ref{fig:lambda_B1_L20} where even less beams are selected (more elements in $\lambdav$ turn out to be zero) because feedback size is extremely small and the DNN chooses accordingly to train effective channels with fewer coefficients.
\section{Conclusion}
We proposed a DNN-based channel training and precoding scheme with channel statistics knowledge at the BS, for FDD massive MIMO systems. The DNN is trained for an ensemble of channel statistics (provided by the cell geometric environment), and generates, for any given input of user channel statistics a pre-beamforming matrix that maps original channels to effective channels such that the DL sum-rate is maximized. The proposed system works with analog feedback and requires no DNN implemented at the user side, which makes it far more practical than most DNN-based approaches in the literature. Our numerical results showed the significant advantage offered by this architecture for both sparse and rich scattering scenarios and various feedback sizes.

\newcommand{\plotwidth}{0.3}

\begin{figure*}[t] 
	\centering
	\begin{subfigure}[b]{\plotwidth\textwidth}
		\includegraphics[trim={0 5 0 0},width=\textwidth]{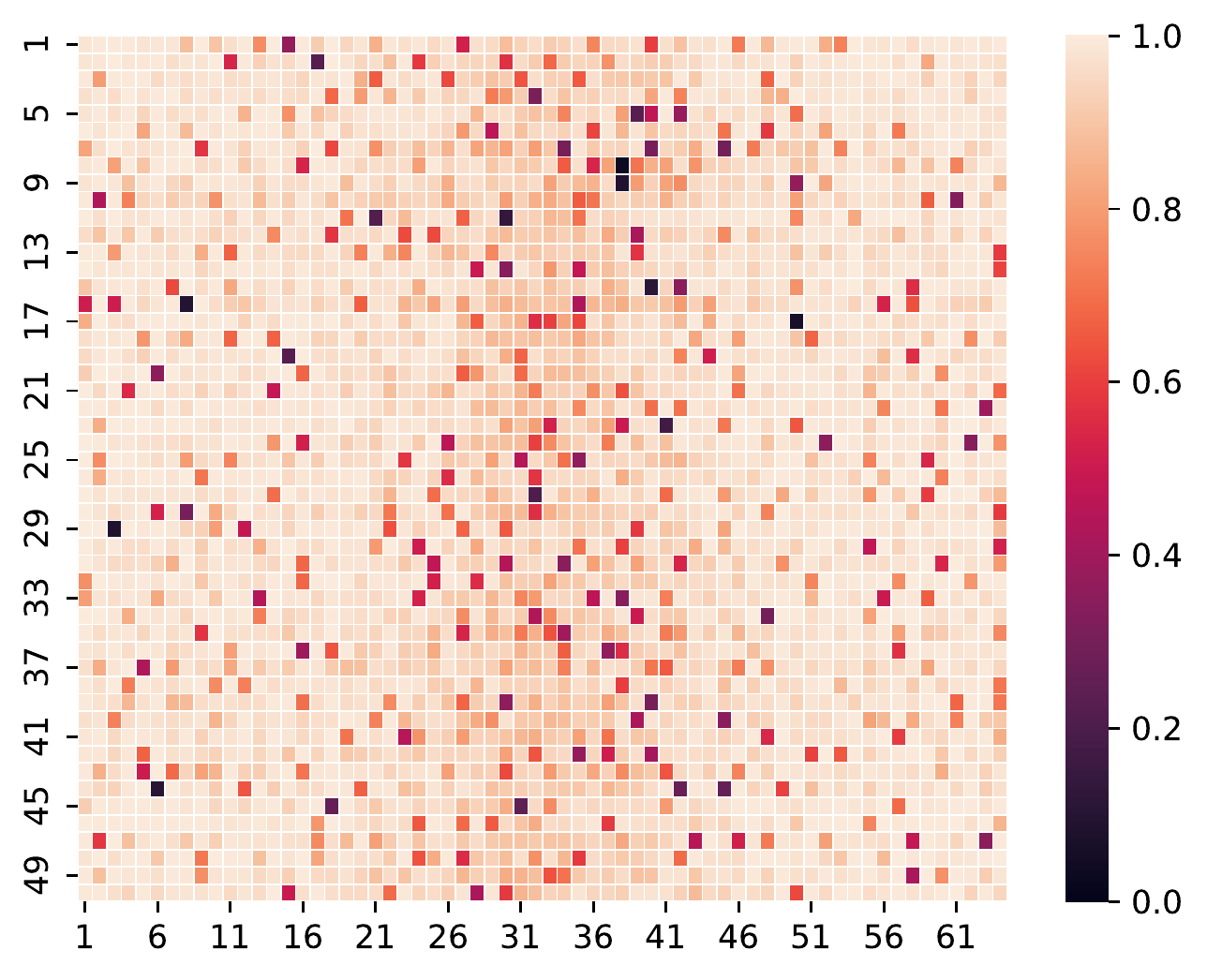}
		\caption{$L=2, B=1$}
		\label{fig:lambda_B1_L2}
		\vspace{-1mm}
	\end{subfigure}
	~ 
	~ 
	\begin{subfigure}[b]{\plotwidth\textwidth}
		\includegraphics[trim={0 5 0 0},width=\textwidth]{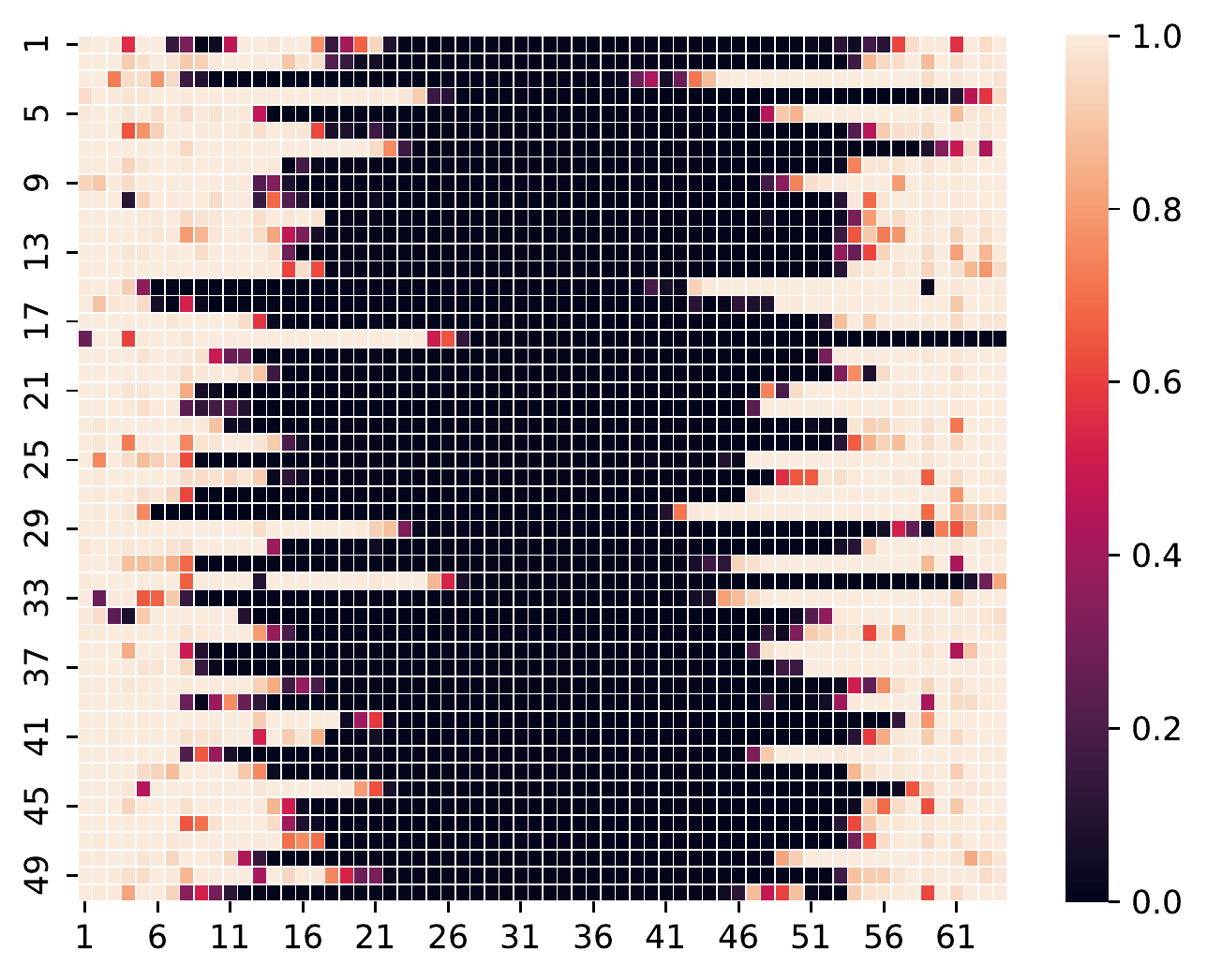}
		\caption{$L=20, B=40$}
		\label{fig:lambda_B40_L20}
		\vspace{-1mm}
	\end{subfigure}
		~
	\begin{subfigure}[b]{\plotwidth\textwidth}
		\includegraphics[trim={0 5 0 0},width=\textwidth]{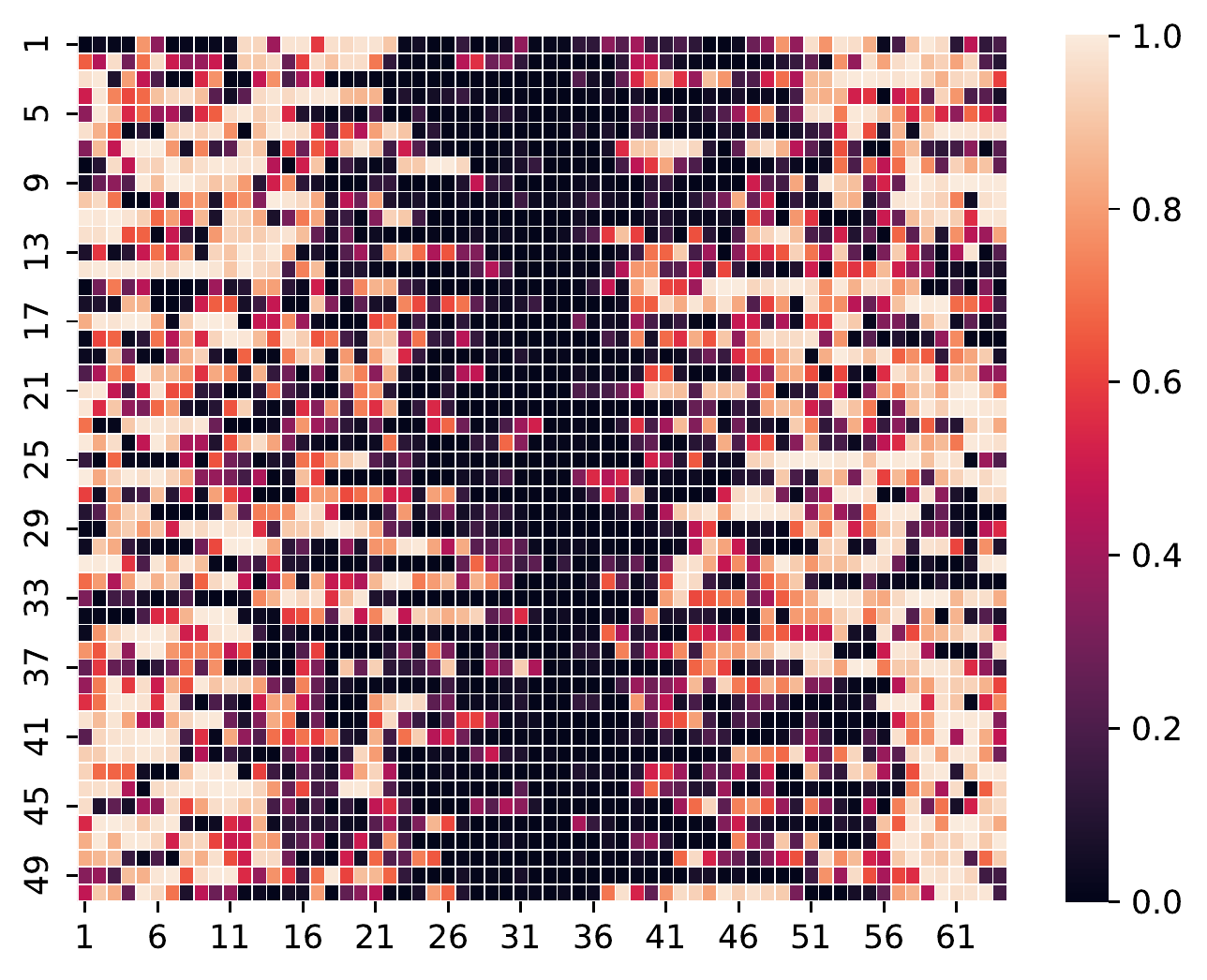}
		\caption{$L=20, B=1$}
		\label{fig:lambda_B1_L20}
		\vspace{-1mm}
	\end{subfigure}
	\caption{Optimized $\lambdav$ instances of the proposed DNN for 50 covariance realizations (stacked as rows)}
	\label{fig:lambda}
	\vspace{-6mm}
\end{figure*}

{\small
	\bibliographystyle{IEEEtran}
	\bibliography{references}

\begin{thebibliography}{10}
\providecommand{\url}[1]{#1}
\csname url@samestyle\endcsname
\providecommand{\newblock}{\relax}
\providecommand{\bibinfo}[2]{#2}
\providecommand{\BIBentrySTDinterwordspacing}{\spaceskip=0pt\relax}
\providecommand{\BIBentryALTinterwordstretchfactor}{4}
\providecommand{\BIBentryALTinterwordspacing}{\spaceskip=\fontdimen2\font plus
\BIBentryALTinterwordstretchfactor\fontdimen3\font minus
  \fontdimen4\font\relax}
\providecommand{\BIBforeignlanguage}[2]{{%
\expandafter\ifx\csname l@#1\endcsname\relax
\typeout{** WARNING: IEEEtran.bst: No hyphenation pattern has been}%
\typeout{** loaded for the language `#1'. Using the pattern for}%
\typeout{** the default language instead.}%
\else
\language=\csname l@#1\endcsname
\fi
#2}}
\providecommand{\BIBdecl}{\relax}
\BIBdecl

\bibitem{eisen2020optimal}
M.~Eisen and A.~Ribeiro, ``Optimal wireless resource allocation with random
  edge graph neural networks,'' \emph{IEEE Transactions on Signal Processing},
  vol.~68, pp. 2977--2991, 2020.

\bibitem{cui2019spatial}
W.~Cui, K.~Shen, and W.~Yu, ``Spatial deep learning for wireless scheduling,''
  \emph{IEEE Journal on Selected Areas in Communications}, vol.~37, no.~6, pp.
  1248--1261, 2019.

\bibitem{balevi2020massive}
E.~Balevi, A.~Doshi, and J.~G. Andrews, ``Massive {MIMO} channel estimation
  with an untrained deep neural network,'' \emph{IEEE Transactions on Wireless
  Communications}, vol.~19, no.~3, pp. 2079--2090, 2020.

\bibitem{mashhadi2021pruning}
M.~B. Mashhadi and D.~G{\"u}nd{\"u}z, ``Pruning the pilots: Deep learning-based
  pilot design and channel estimation for {MIMO-OFDM} systems,'' \emph{IEEE
  Transactions on Wireless Communications}, vol.~20, no.~10, pp. 6315--6328,
  2021.

\bibitem{hojatian2021unsupervised}
H.~Hojatian, J.~Nadal, J.-F. Frigon, and F.~Leduc-Primeau, ``Unsupervised deep
  learning for massive {MIMO} hybrid beamforming,'' \emph{IEEE Transactions on
  Wireless Communications}, vol.~20, no.~11, pp. 7086--7099, 2021.

\bibitem{honkala2021deeprx}
M.~Honkala, D.~Korpi, and J.~M. Huttunen, ``Deeprx: Fully convolutional deep
  learning receiver,'' \emph{IEEE Transactions on Wireless Communications},
  vol.~20, no.~6, pp. 3925--3940, 2021.

\bibitem{marzetta2006fast}
T.~L. Marzetta and B.~M. Hochwald, ``Fast transfer of channel state information
  in wireless systems,'' \emph{IEEE Transactions on Signal Processing},
  vol.~54, no.~4, pp. 1268--1278, 2006.

\bibitem{alrabeiah2019deep}
M.~Alrabeiah and A.~Alkhateeb, ``Deep learning for {TDD} and {FDD} massive
  {MIMO}: Mapping channels in space and frequency,'' in \emph{2019 53rd
  asilomar conference on signals, systems, and computers}.\hskip 1em plus 0.5em
  minus 0.4em\relax IEEE, 2019, pp. 1465--1470.

\bibitem{arnold2019towards}
M.~Arnold, S.~D{\"o}rner, S.~Cammerer, J.~Hoydis, and S.~ten Brink, ``Towards
  practical {FDD} massive {MIMO}: {CSI} extrapolation driven by deep learning
  and actual channel measurements,'' in \emph{2019 53rd Asilomar Conference on
  Signals, Systems, and Computers}.\hskip 1em plus 0.5em minus 0.4em\relax
  IEEE, 2019, pp. 1972--1976.

\bibitem{yang2019deep}
Y.~Yang, F.~Gao, G.~Y. Li, and M.~Jian, ``Deep learning-based downlink channel
  prediction for {FDD} massive {MIMO} system,'' \emph{IEEE Communications
  Letters}, vol.~23, no.~11, pp. 1994--1998, 2019.

\bibitem{ma2020data}
X.~Ma and Z.~Gao, ``Data-driven deep learning to design pilot and channel
  estimator for massive {MIMO},'' \emph{IEEE Transactions on Vehicular
  Technology}, vol.~69, no.~5, pp. 5677--5682, 2020.

\bibitem{mashhadi2020distributed}
M.~B. Mashhadi, Q.~Yang, and D.~G{\"u}nd{\"u}z, ``Distributed deep
  convolutional compression for massive {MIMO} {CSI} feedback,'' \emph{IEEE
  Transactions on Wireless Communications}, vol.~20, no.~4, pp. 2621--2633,
  2020.

\bibitem{wen2018deep}
C.-K. Wen, W.-T. Shih, and S.~Jin, ``Deep learning for massive {MIMO} {CSI}
  feedback,'' \emph{IEEE Wireless Communications Letters}, vol.~7, no.~5, pp.
  748--751, 2018.

\bibitem{guo2020deep}
J.~Guo, C.-K. Wen, and S.~Jin, ``Deep learning-based {CSI} feedback for
  beamforming in single-and multi-cell massive {MIMO} systems,'' \emph{IEEE
  Journal on Selected Areas in Communications}, vol.~39, no.~7, pp. 1872--1884,
  2020.

\bibitem{sohrabi2021deep}
F.~Sohrabi, K.~M. Attiah, and W.~Yu, ``Deep learning for distributed channel
  feedback and multiuser precoding in {FDD} massive {MIMO},'' \emph{IEEE
  Transactions on Wireless Communications}, vol.~20, no.~7, pp. 4044--4057,
  2021.

\bibitem{xie2016unified}
H.~Xie, F.~Gao, S.~Zhang, and S.~Jin, ``A unified transmission strategy for
  {TDD/FDD} massive {MIMO} systems with spatial basis expansion model,''
  \emph{IEEE Transactions on Vehicular Technology}, vol.~66, no.~4, pp.
  3170--3184, 2016.

\bibitem{miretti2018fdd}
L.~Miretti, R.~L.~G. Cavalcante, and S.~Stanczak, ``{FDD} massive {MIMO}
  channel spatial covariance conversion using projection methods,'' in
  \emph{2018 IEEE International Conference on Acoustics, Speech and Signal
  Processing (ICASSP)}.\hskip 1em plus 0.5em minus 0.4em\relax IEEE, 2018, pp.
  3609--3613.

\bibitem{haghighatshoar2018multi}
S.~Haghighatshoar, M.~B. Khalilsarai, and G.~Caire, ``Multi-band covariance
  interpolation with applications in massive {MIMO},'' in \emph{2018 IEEE
  International Symposium on Information Theory (ISIT)}.\hskip 1em plus 0.5em
  minus 0.4em\relax IEEE, 2018, pp. 386--390.

\bibitem{khalilsarai2018fdd}
M.~B. Khalilsarai, S.~Haghighatshoar, X.~Yi, and G.~Caire, ``{FDD} massive
  {MIMO} via {UL/DL} channel covariance extrapolation and active channel
  sparsification,'' \emph{IEEE Transactions on Wireless Communications},
  vol.~18, no.~1, pp. 121--135, 2018.

\bibitem{caire2010multiuser}
G.~Caire, N.~Jindal, M.~Kobayashi, and N.~Ravindran, ``Multiuser {MIMO}
  achievable rates with downlink training and channel state feedback,''
  \emph{IEEE Transactions on Information Theory}, vol.~56, no.~6, pp.
  2845--2866, 2010.

\bibitem{lu2020multi}
Z.~Lu, J.~Wang, and J.~Song, ``Multi-resolution {CSI} feedback with deep
  learning in massive {MIMO} system,'' in \emph{ICC 2020-2020 IEEE
  International Conference on Communications (ICC)}.\hskip 1em plus 0.5em minus
  0.4em\relax IEEE, 2020, pp. 1--6.

\bibitem{caire2018ergodic}
G.~Caire, ``On the ergodic rate lower bounds with applications to massive
  {MIMO},'' \emph{IEEE Transactions on Wireless Communications}, vol.~17,
  no.~5, pp. 3258--3268, 2018.

\bibitem{adhikary2013joint}
A.~Adhikary, J.~Nam, J.-Y. Ahn, and G.~Caire, ``Joint spatial division and
  multiplexing—the large-scale array regime,'' \emph{IEEE Transactions on
  Information Theory}, vol.~59, no.~10, pp. 6441--6463, 2013.

\bibitem{zhu2017asymptotic}
Z.~Zhu and M.~B. Wakin, ``On the asymptotic equivalence of circulant and
  toeplitz matrices,'' \emph{IEEE Transactions on Information Theory}, vol.~63,
  no.~5, pp. 2975--2992, 2017.

\bibitem{paszke2019pytorch}
A.~Paszke, S.~Gross, F.~Massa, A.~Lerer, J.~Bradbury, G.~Chanan, T.~Killeen,
  Z.~Lin, N.~Gimelshein, L.~Antiga \emph{et~al.}, ``Pytorch: An imperative
  style, high-performance deep learning library,'' \emph{Advances in Neural
  Information Processing Systems}, 2019.

\bibitem{kingma2015adam}
D.~P. Kingma and B.~Jimmy, ``Adam: {A} method for stochastic optimization,'' in
  \emph{3rd International Conference on Learning Representations, {ICLR}},
  2015.

\bibitem{ioffe2015BN}
S.~Ioffe and C.~Szegedy, ``Batch normalization: Accelerating deep network
  training by reducing internal covariate shift,'' in \emph{Proceedings of the
  32nd International Conference on Machine Learning, {ICML}}, 2015, p.
  448–456.

\bibitem{liu2020statistical}
H.~Liu, X.~Yuan, and Y.~J. Zhang, ``Statistical beamforming for {FDD} downlink
  massive {MIMO} via spatial information extraction and beam selection,''
  \emph{IEEE Transactions on Wireless Communications}, vol.~19, no.~7, pp.
  4617--4631, 2020.

\end{thebibliography}
}

\end{document}